\documentclass[conference]{IEEEtran}
\usepackage{cite}
\usepackage{amsmath,amssymb,amsfonts}
\usepackage{algorithmic}
\usepackage{graphicx}
\usepackage{textcomp}
\usepackage{xcolor}
\usepackage{comment}
\usepackage{enumitem}
\usepackage{graphicx}
\usepackage{subcaption}
\usepackage{tabularx}
\def\BibTeX{{\rm B\kern-.05em{\sc i\kern-.025em b}\kern-.08em
    T\kern-.1667em\lower.7ex\hbox{E}\kern-.125emX}}
\title{Perception-Aware Communication Middleware for Distributed Visual Perception in UAV Swarms\\
\vspace{-0.2in}
}

\author{
\IEEEauthorblockN{
Manveen Kaur\IEEEauthorrefmark{1}
Kevin Loi\IEEEauthorrefmark{1},
Ifunanya Okafor\IEEEauthorrefmark{1},
Daniel Ng\IEEEauthorrefmark{1},
Joseph Lucey-Renteria\IEEEauthorrefmark{2}
}

\IEEEauthorblockA{
\IEEEauthorrefmark{1}
Department of Computer Science, California State University, Los Angeles\\
Los Angeles, CA, USA\\
\{mkaur39, kloi2, iokafor6, dng14\}@calstatela.edu
}

\IEEEauthorblockA{
\IEEEauthorrefmark{2}
Department of Civil Engineering, California State University, Los Angeles\\
Los Angeles, CA, USA\\
jlucey2@calstatela.edu
}

\thanks{This work was supported by the Provost SEED Grant at California State University, Los Angeles.}
\thanks{Manveen Kaur and Kevin Loi contributed equally to this work. Manveen Kaur is the corresponding author (email: mkaur39@calstatela.edu).}
}
\IEEEoverridecommandlockouts
\begin{document}

\maketitle
\vspace{-0.3in}

\begin{abstract}
Unmanned Aerial Vehicle (UAV) swarms increasingly support safety-critical applications that rely on distributed visual perception. Meeting the low-latency requirements of these applications can require perception models to execute within the swarm on inference-capable UAVs, creating a need for efficient UAV-to-UAV transport of high-bandwidth perception data. However, the Quality-of-Service (QoS) requirements of perception differ from conventional packet-level QoS; successful delivery of individual packets does not ensure that a complete, timely, and usable image is available for inference. We present a novel perception-aware communication middleware that treats complete perception-data samples as the communication objects for which QoS must be satisfied. The middleware extends a lightweight UDP broker-based publish-subscribe architecture with perception-specific services, including image fragmentation and reconstruction, concurrent packet transmission, priority-aware scheduling, and image quality assessment. The middleware is evaluated on a heterogeneous hardware testbed emulating a UAV swarm using YOLOv8n object detection. Experimental results demonstrate low end-to-end application latency, substantially higher throughput than a lightweight UDP broker, effective prioritization of perception traffic under increasing background load, and mitigation of object-detection degradation through middleware-level image quality assessment. This work provides an initial framework for integrating AI-specific data handling into communication middleware to support emerging distributed AI applications in multi-agent mobile cyber-physical systems.

\end{abstract}


\section{Introduction}
Unmanned Aerial Vehicle (UAV) swarms, defined as collaborative and cooperative group of UAVs working to meet a common objective, can be utilized to support numerous safety-critical and time-sensitive applications in critical infrastructure domains. Many of these applications, e.g. post-disaster search and rescue (SAR) and battlefield reconnaissance, are driven by continuous video sensing to provide real-time situational awareness~\cite{zhao2025active}. These applications increasingly rely on compute- and memory- intensive deep learning-based perception models, such as object detection, to achieve mission objectives~\cite{moon2023swarm,horyna2023decentralized}. Further, safety-critical applications must also operate under stringent Quality of Service (QoS) guarantees for latency and reliability, and in dynamic environments that frequently lack reliable communication infrastructure or low-latency access to remote computational resources~\cite{moon2023swarm}.

Because UAV swarms are typically heterogeneous, with most UAVs constrained in compute, memory, and communication resources~\cite{alqudsi2025uav}, perception applications may require distributed operation where image and video sensing are decoupled from perception-model execution. Furthermore, image and video sensing generate high-bandwidth perception data whose timely and reliable delivery directly influences the perception models' inference accuracy and overall application performance. Therefore, UAV swarms must support the transport and processing of high-bandwidth perception data while being mindful of strict QoS requirements.  Conventional approaches for incorporating perception models in UAV swarms involve running model inference on remote cloud or edge computing infrastructure~\cite{alqudsi2025uav}. However, UAV swarms deployed in disaster response, environmental monitoring, or battlefield operations cannot always rely on persistent or timely connectivity to such infrastructure.

Integrating Edge AI~\cite{singh2023edge}, i.e., running AI models on local devices, with Mobile Edge Computing, i.e., utilizing computationally capable UAVs as mobile edge nodes, enables near-real-time onboard perception by executing perception models locally within the swarm. Here, UAVs can be designated as \textit{Edge AI UAVs} or \textit{Sensing UAVs}, with UAV-to-UAV communication supporting perception-data exchange between the two. Three technological drivers underpin this vision: \textbf{(1)} advances in compact AI accelerator-equipped System-on-Chips (SoCs) that enable Edge AI~\cite{reuther2022ai}, \textbf{(2)} the design of lightweight and distributed AI/ML models for resource-constrained multi-agent systems~\cite{su2022ai,ding2023distributed}, and \textbf{(3)} the development of \textit{perception-aware communication middleware} that extends QoS support beyond individual packets to complete data samples required for inference. Unlike conventional packet-centric communication, this middleware should identify, reconstruct, and prioritize complete perception inputs to meet application-level QoS requirements. This distinction is critical because successful delivery of individual packets does not guarantee usable perception data; fragments from multiple incomplete images may satisfy packet-level QoS requirements while providing no complete image suitable for inference. While recent research has demonstrated substantial progress toward the first two drivers through the emergence of numerous AI accelerator-equipped embedded platforms~\cite{qualcomm_iq8_brief,nvidia_jetson_orin_brief} and model optimization techniques~\cite{su2022ai}, distributed perception applications remain fundamentally dependent on the timely and reliable transport of high-bandwidth perception data among UAVs. Existing lightweight communication middleware, including MQTT, CoAP, and ZeroMQ, provide efficient message transport but remain unaware of data samples and the needs of perception models~\cite{naik2017choice,liu2025performance}. 

In this work, we present a perception-aware communication middleware for UAV swarms that treats complete image samples as the data units for which QoS requirements must be satisfied. The middleware extends a UDP broker-based publish-subscribe architecture with services tailored to distributed perception and transmission of complete images, including thread pooling for high-throughput transmission of image fragments, image-specific header fields to maintain semantic context over multiple image fragments, priority-aware scheduling, middleware-level image reconstruction, and image quality assessment to filter degraded inputs before inference. The middleware is implemented in C as an application layer service, and evaluated on a heterogeneous multi-agent hardware testbed representative of resource-constrained UAV swarms. Experimental results demonstrate high-throughput perception-data transport with low end-to-end application latency, effective prioritization of QoS-sensitive image traffic under increasing network load, and filtering of degraded images to preserve downstream object-detection performance.

The remainder of this paper is organized as follows. Section II presents the background and related work. Section III introduces the reference UAV swarm architecture and describes the salient features of the proposed communication middleware. Section IV details the experimental methodology and evaluation strategy. Section V presents and discusses the experimental results. Finally, Section VI concludes the paper.

\section{Background and Related Works} \label{background}

Various technological advances are enabling distributed perception in resource-constrained UAV swarms. Advances in embedded edge AI hardware have substantially increased onboard computational capability, with modern System-on-Chips (SoCs) integrating dedicated AI accelerators, including Graphics Processing Units (GPUs), application-specific integrated circuits (ASICs), and field-programmable gate arrays (FPGAs), to accelerate deep learning inference. Examples include the Qualcomm DragonWing~\cite{qualcomm_iq8_brief} and NVIDIA Jetson families~\cite{nvidia_jetson_orin_brief}. In parallel, model optimization techniques such as pruning, quantization, and knowledge distillation reduce model size, memory consumption, and inference latency~\cite{su2022ai}, while inference frameworks such as NVIDIA TensorRT and TensorFlow Lite further optimize execution on embedded hardware. Distributed AI paradigms, including Federated Learning and Multi-Agent Reinforcement Learning, further enable AI workloads to be distributed across multiple agents~\cite{su2022ai,ding2023distributed}.

Deep learning-based visual perception has consequently become a fundamental capability for safety-critical autonomous UAV swarm applications. In SAR, representative approaches include using RGB cameras and a CNN-based object detector to identify survivors~\cite{horyna2023decentralized} and executing a pruned YOLOv5 model onboard each UAV for object detection~\cite{moon2023swarm}. However, many existing perception approaches assume either sufficient computational resources for local inference or reliable, low-latency, high-bandwidth connectivity to remote edge infrastructure. While some works explicitly consider communication constraints in distributed UAV perception, primarily by reducing communication overhead through selective information exchange, optimizing communication resources, or improving network connectivity~\cite{hu2022where2comm,mehrabi2025adaptive}, comparatively less attention has been given to communication middleware that manages perception data itself and its end-to-end transport between sensing and edge AI UAVs.

Existing communication middleware, including MQTT, CoAP, and ZeroMQ, provides efficient and lightweight message transport for distributed systems but is largely application agnostic. MQTT provides broker-based publish-subscribe messaging with multiple QoS levels; CoAP provides lightweight RESTful communication, typically over UDP, for constrained devices; and ZeroMQ provides high-performance asynchronous messaging patterns for distributed applications~\cite{naik2017choice,liu2025performance}. While suitable for general-purpose communication, these technologies do not natively manage perception data as complete communication units, including their fragmentation and reconstruction or preservation of application-level priority across image fragments. Emerging technologies such as Multi-access Edge Computing (MEC) and 5G NR Sidelink provide complementary capabilities for low-latency edge computing and direct device-to-device communication, but depend on compatible network infrastructure and hardware~\cite{bagheri20215g}. Thus, a gap remains for \textit{perception-aware communication middleware} that bridges communication and perception by managing complete perception-data units while satisfying the latency, reliability, and bandwidth requirements of distributed edge inference.

\section{Solution Design} \label{design}
\begin{figure}[ht]
    \centering
    \includegraphics[width=1\linewidth]{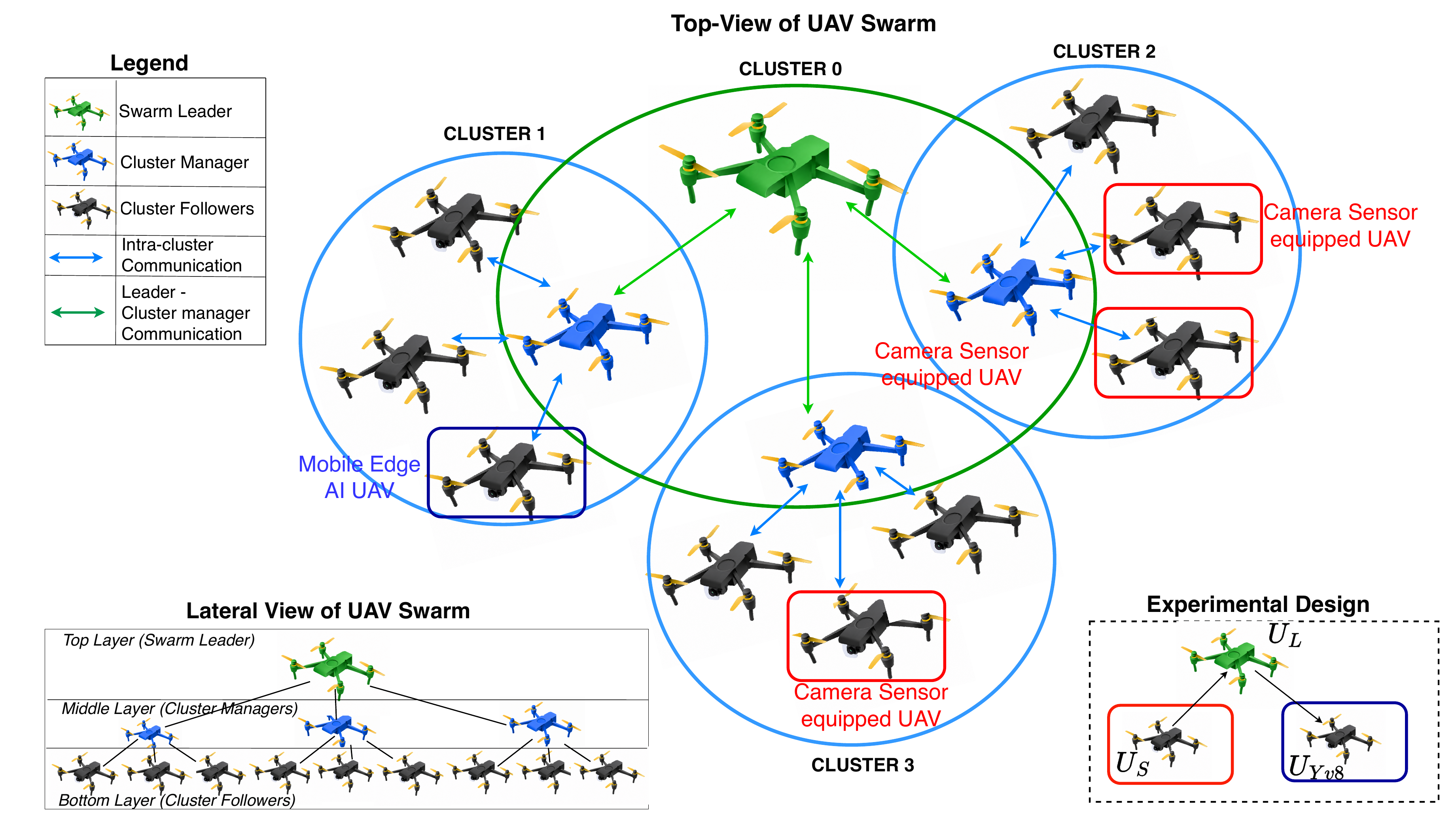}
    \caption{Proposed UAV Swarm Configuration and Experimental Testbed Configuration (bottom right)}
    \label{fig:F1}
\end{figure}

\textbf{Swarm Configuration:} The proposed communication middleware assumes a distributed, decentralized, hierarchical UAV swarm architecture that supports scalable coordination among heterogeneous UAVs operating in communication-constrained environments~\cite{kaur2023design}. As shown in Fig.~\ref{fig:F1}, the swarm is partitioned into multiple heterogeneous clusters, each managed by a cluster manager responsible for coordinating local cluster operations. A swarm leader oversees inter-cluster coordination by forming a top-level cluster with the cluster managers,  facilitating information exchange among clusters.
Tasks are assigned according to UAV capabilities. Resource-constrained follower UAVs primarily perform sensing and data collection, while computationally capable cluster managers and the swarm leader coordinate task execution. Depending on mission requirements, one or more computationally capable UAVs within the swarm may serve as mobile edge AI nodes, executing perception models on behalf of resource-constrained sensing UAVs. To improve robustness and eliminate a single point of failure, the roles of swarm coordination and edge AI execution are decoupled; thus, swarm leaders may not function as mobile edge AI nodes. It is not necessary for each cluster to have its own edge AI node(s). While the proposed middleware can support all inter-UAV communication, including control, coordination, and application data exchange, the remainder of this paper focuses exclusively on the communication supporting distributed visual perception.

\textbf{Proposed Communication Middleware:} The proposed middleware extends a lightweight UDP broker-based publish-subscribe architecture and comprises two software modules: a \textit{client module}, which provides publisher and subscriber functions on follower and edge AI UAVs, and a \textit{broker module}, which executes on cluster managers and swarm leaders to facilitate communication among clients. The swarm hierarchy determines the broker servicing each client, while inter-broker communication is supported through broker-to-broker subscriptions. As the underlying publish-subscribe mechanisms, including topic-based publication, subscription, and broker operation, are well established, we refer readers to prior literature for further details. The middleware supports two communication modes: \textit{best-effort}, in which messages are transmitted without acknowledgment, and \textit{reliable}, in which published messages require acknowledgment, analogous to MQTT QoS 0 and QoS 1, respectively. This work evaluates the best-effort mode.

Unlike conventional packet-centric communication, the middleware treats perception data as the fundamental communication object and provides the following core services:


\begin{itemize}[leftmargin=0pt]
    \item \textbf{\textit{Image Fragmentation and Reconstruction:}} Standard-resolution images exceed the maximum payload supported by a network packet, requiring fragmentation before transmission. The client code fragments each sensing image, and adds semantic context through header fields, such as \(\mathbf{image\_id}\), \(\mathbf{fragment\_id}\), and \(\mathbf{total\_fragments}\). The client code computes the total fragments,  \(\mathbf{F}\), in an image as \(\mathbf{S/P}\), where \(\mathbf{S}\) is the file size and \(\mathbf{P}\) is the maximum payload size for each packet. Each fragment is assigned a \(\mathbf{msg\_id}\), \(\mathbf{image\_id}\), and \(\mathbf{fragment\_id}\) prior to transmission. Zero-padding is used if a fragment size is \(\leq\mathbf{P}\). Upon receiving image fragments, the image reconstruction involves writing each fragment into a new file in sequence (using \(\mathbf{fragment\_id}\)). Image fragmentation and reconstruction preserve the image file format. At the receiver, the client module maintains an image log that tracks the reconstruction status of each partially received image. Once all fragments of an image are received and reconstruction is complete, its entry is removed from the log. The client waits for a predefined timeout period for all fragments to arrive; if the timeout expires, the image is reconstructed using the available fragments. The timeout is dynamically determined using an Exponentially Weighted Moving Average (EWMA) of the measured round-trip time (RTT).
    
    \item \textbf{\textit{Concurrent Packet Transmission:}} Conventional sequential processing and transmission of image fragments can limit achievable throughput and increase the time required to deliver a complete image to the perception model. To improve throughput and reduce image delivery latency, the proposed middleware employs concurrent packet transmission using a fixed-size thread pool in both client and broker modules. The thread pool supports two task types, image fragmentation and packet transmission, submitted to a shared job queue. Fragmentation tasks divide an image concurrently by assigning distinct image regions to available worker threads; once fragmentation is complete, transmission tasks publish the resulting fragments through UDP sockets to the broker or subscriber. Mutex-protected critical sections ensure thread-safe access to shared buffers. The number of worker threads is configurable, balancing communication performance against processor utilization; larger pools increase concurrency and can reduce image delivery latency at the cost of greater CPU utilization, while smaller pools reduce computational overhead but may increase delivery time.

    \item \textbf{\textit{Priority-Aware Packet Scheduling:}} A client assigns a priority level to every transmitted packet, and the broker schedules outgoing messages using a multi-level circular ring-buffer priority queue. The scheduler maintains multiple independent circular buffers, one for each priority level. Incoming packets are inserted into the tail of the buffer corresponding to their assigned priority, while transmission always services the highest non-empty priority queue first.  A limitation of this priority scheduling is the possibility of starvation for lower-priority traffic during sustained bursts of high-priority image transmission. To mitigate this effect, the middleware employs priority aging. Each packet maintains a starvation counter that increases while waiting in the queue. When the counter exceeds a predefined threshold determined by application needs, the packet's priority is increased, ensuring that lower-priority packets are also serviced without significantly affecting latency for high-priority perception traffic.

    \item \textbf{\textit{Image Quality Assessment (IQA):}} Various factors can degrade the quality of received perception data and reduce the performance of the perception model. Our current IQA implementation focuses on image degradation due to noisy sensing data. To prevent degraded images from being forwarded to the perception model, the proposed middleware performs IQA following image reconstruction using a computationally efficient sharpness metric called the Variance of the Laplacian which we implement natively in the middleware~\cite{pertuz2013analysis,csimcsek2021novel}. The reconstructed image is converted to grayscale and processed using a standard four-neighbor discrete Laplacian operator to identify edges. The variance of the Laplacian response provides a sharpness score: high variance indicates strong edge information and a sharper image, while low variance indicates blur or degradation that may reduce inference reliability. The computed variance is compared against a configurable threshold selected through empirical calibration. Images whose sharpness score falls below the threshold are classified as low quality and are flagged. The current version of the middleware discards these images; however, it is possible to flag them for review by the perception models.  
\end{itemize}

\section{Experiment Strategy} \label{exp-strategy}
We consider an exemplar application of a heterogeneous UAV swarm supporting a latency-sensitive reconnaissance mission on the battlefield. The swarm is tasked with surveying an area of interest, and uses video sensing and a YOLOv8n perception model to identify objects of interest. Simplifying for validation, we use the swarm configuration shown in the bottom-right insert in Fig.~\ref{fig:F1}. Here, one camera-sensor equipped follower UAV \((U_S)\) conducts video-sensing, and a pre-trained YOLOv8n model runs on the Mobile Edge AI UAV \((U_{Yv8})\). The client software runs on \(U_S\) and \(U_{Yv8}\), while the broker software runs on the swarm leader \((U_L)\). Communication between \(U_S\) and \(U_{Yv8}\) follows the path \(U_S \leftrightarrow U_L \leftrightarrow U_{Yv8}\). We assume that the outcome of the perception model is utilized as an input for future path planning of the swarm. Although more recent YOLO variants offer improved computational efficiency, YOLOv8n was selected as the mature and stable release available during middleware development. Since the proposed middleware is independent of the underlying perception model, the choice of detector does not affect its communication mechanisms or the conclusions of this work. We train YOLOv8n using the AU-AIR dataset~\cite{bozcan2020air}, which contains over 32,000 annotated RGB image frames extracted from low-altitude UAV videos across eight traffic-related object classes. The experimental perception task focuses on \textit{car} detection. Model training was performed on a computationally-capable machine, and the pre-trained model was deployed on \(U_{Yv8}\). 

\textbf{Hardware Testbed:} The proposed communication middleware was evaluated using a heterogeneous hardware testbed emulating the experimental configuration in Fig.~\ref{fig:F1}. A Raspberry Pi 4 (RPi4), Intel Next Unit of Computing (NUC), and NVIDIA Jetson AGX Orin (Orin) serve as \(U_S\), \(U_L\), and \(U_{Yv8}\), respectively. The RPi4 represents a resource-constrained UAV, while the NUC represents a comparatively more computationally capable UAV. The Orin serves as the edge AI node, providing dedicated hardware for neural-network inference not available on the RPi4 or NUC. All nodes are interconnected via wired Ethernet operating at 100 Mbps to isolate middleware performance from wireless effects such as channel contention, scheduling, and packet collisions, as well as bandwidth bottlenecks. Similarly, node mobility is not considered in this setup to avoid complications arising from network disruptions. System clocks are synchronized using the Chrony implementation of the Network Time Protocol (NTP). 

\textbf{Experimental Design:} 
We conducted three experiments. \textit{Experiment 1} transmits a set of 182 images from the AU-AIR dataset (unseen during YOLOv8n training) from \(U_S\) to \(U_{Yv8}\) to evaluate network-level communication, middleware, and application performance. We measure \textit{end-to-end communication latency} as the time between transmission and reception of each image fragment; \textit{middleware latency} as the time between each fragment's transmission until completion of image reconstruction and IQA; and \textit{application latency} as the time between transmission of the first image fragment and completion of YOLOv8n inference. Corresponding packet-, middleware-, and application-level throughput are also measured.

\textit{Experiment 2} evaluates the middleware's priority-aware packet scheduling under increasing network load. Alongside transmitting the 182-image dataset from \(U_S\) to \(U_{Yv8}\), a second sender (\(U'_S\)) simultaneously transmits equivalent volume of data to a second receiver (\(U'_R\)) through the same broker. Image fragments originating from \(U_S\) are assigned a higher priority level (10) than background traffic (priority level 1). Four traffic profiles are evaluated by increasing the transmission rate of \(U'_S\) from \(1\times\) to \(4\times\) that of \(U_S\), and the resulting end-to-end communication latency is measured.

\begin{figure}[h]
    \centering
    \includegraphics[width=1\linewidth]{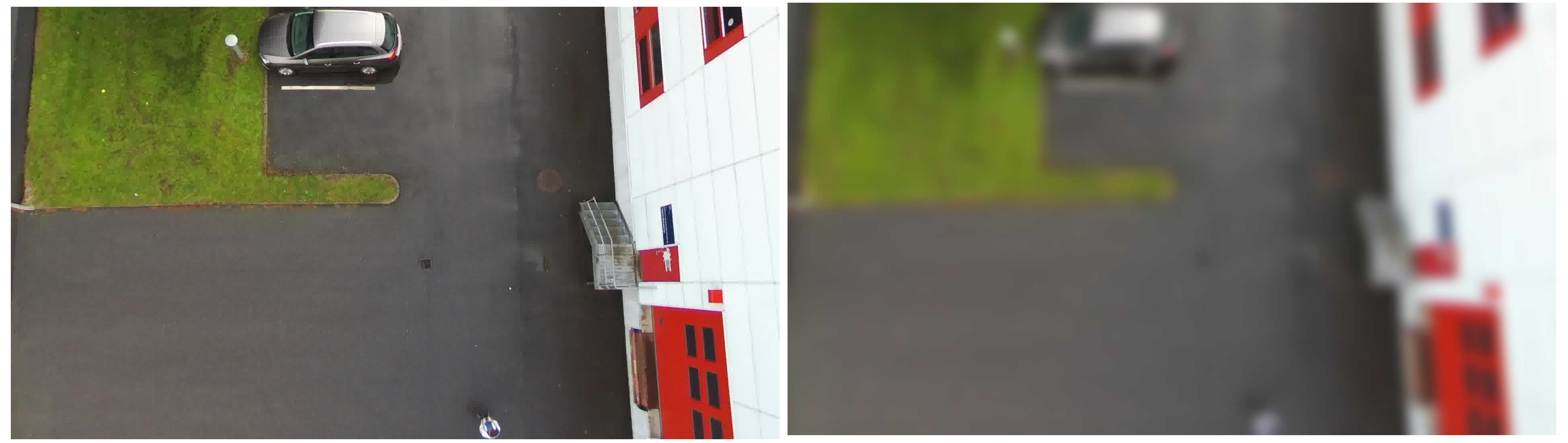}
    \caption{Illustration of a Standard Image and its Degraded Version Created using Gaussian Blur.}
    \label{fig:M1}
\end{figure}

\textit{Experiment 3} evaluates the middleware's IQA feature. To emulate degraded sensing conditions, 20, 40, and 60 of the 182 transmitted images are artificially degraded using Gaussian blur prior to transmission. Gaussian blur reduces image sharpness by averaging neighboring pixel values, with the blur radius controlling the degree of degradation without altering image resolution. Fig.~\ref{fig:M1} provides an illustration of image degradation used. The degraded images are interspersed with unmodified images and transmitted from \(U_S\) to \(U_{Yv8}\). The feature is evaluated on the middleware's ability to filter degraded images and mitigate their impact on YOLOv8n object detection using the standard metrics of precision, recall, mAP@50, and mAP@50--95, whose definitions are omitted for brevity. 

\begin{figure*}[ht]
    \centering

    \begin{subfigure}[t]{0.32\textwidth}
        \centering
        \includegraphics[width=\linewidth, trim=20 10 23 10,clip]{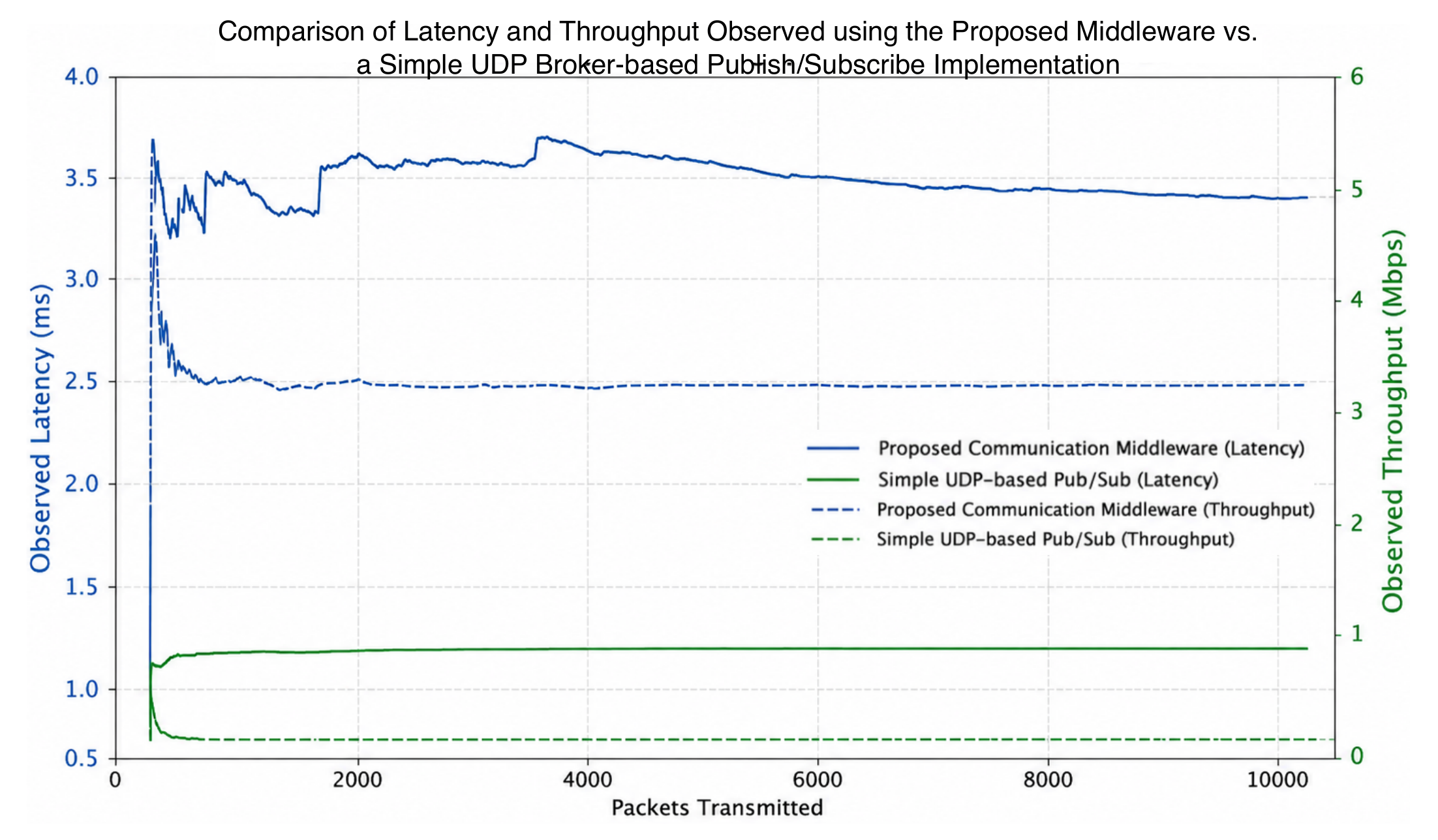}
        \caption{Proposed Middleware vs. a Simple UDP Broker-based Implementation.}
        \label{fig:R1}
    \end{subfigure}
   \hspace{0.0001\textwidth}
    \begin{subfigure}[t]{0.33\textwidth}
        \centering
        \includegraphics[width=\linewidth, trim=25 10 23 10,clip]{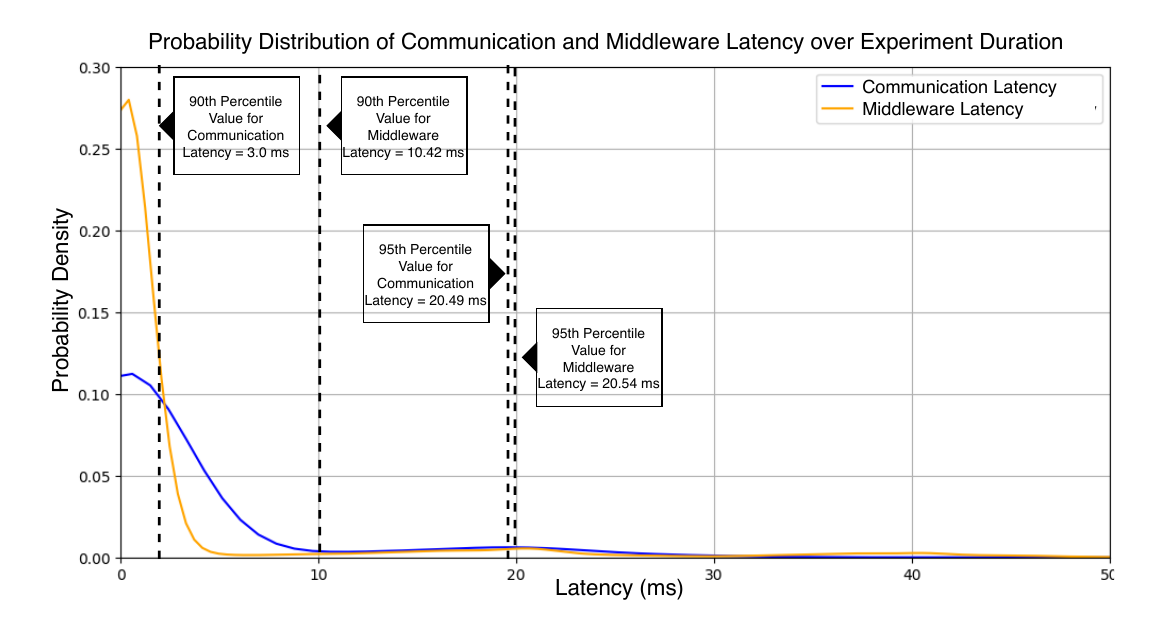}
        \caption{End-to-End Communication and Middleware Latency.}
        \label{fig:R2-1}
    \end{subfigure}
   \hspace{0.005\textwidth}
    \begin{subfigure}[t]{0.29\textwidth}
        \centering
        \includegraphics[width=\linewidth, trim=25 10 23 10,clip]{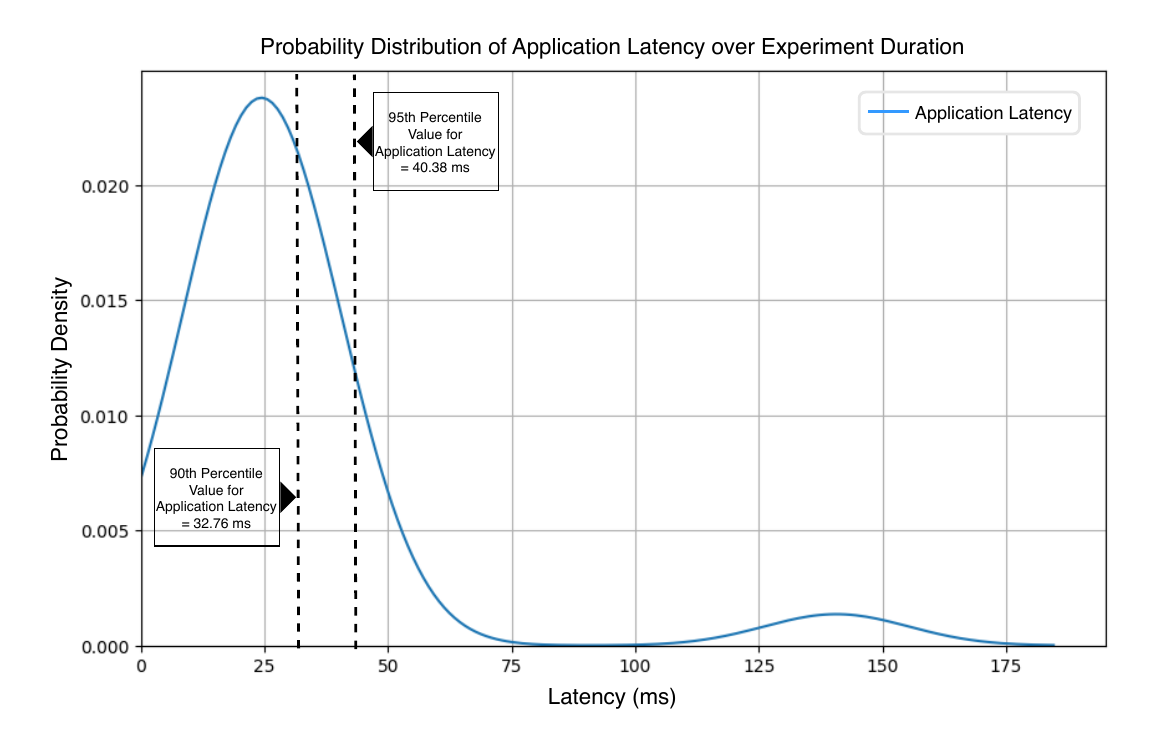}
        \caption{End-to-End Application Latency.}
        \label{fig:R2-2}
    \end{subfigure}

    \caption{Comparison of the Middleware with a Simple UDP broker and Results from Experiment 1.}
    \label{fig:E1}
\end{figure*}

\section{Results and Discussion} \label{results}

Fig.~\ref{fig:R1} compares communication latency and throughput of the proposed middleware against a lightweight UDP broker for transmission of 10,000 packets between a publisher and subscriber in the hardware testbed. The UDP broker implements only basic list-based topic matching, without QoS mechanisms or advanced topic lookup. While the additional functionality and processing introduced by the proposed middleware results in approximately 40\% higher per-packet latency than the UDP broker, threadpooling in the proposed middleware allows multiple packet transmission operations to progress concurrently, increasing the observed throughput by more than 200\%. The proposed middleware trades a smaller increase in individual packet latency for substantially higher aggregate data-transfer throughput, enabling more effective utilization of available communication capacity for high-bandwidth perception data.

Fig.~\ref{fig:R2-1} shows communication and middleware latency observed in Experiment~1. 
The average per-packet communication latency is \(3.175\)~ms, with 90th- and 95th-percentile latencies of \(3.0\)~ms and \(20.49\)~ms, respectively. Thus, 90\% of the 10,000 transmitted image fragments were delivered within \(3.0\)~ms, while the substantially higher 95th-percentile value indicates a small number of significant tail-latency events. The higher tail-end latency values are attributed to the fact that, in the current implementation, multiple worker threads process image fragments and issue transmissions within a short interval, producing synchronized packet bursts that can increase contention at socket and kernel buffers. Thread management will be further optimized in the next iteration of the middleware. The average per-packet middleware latency was \(6.701\)~ms, approximately \(2.1\times\) the average communication latency, with 90th- and 95th-percentile values of \(10.42\)~ms and \(20.54\)~ms, respectively. The increase results from image reconstruction and IQA steps. Ablation experiments with IQA disabled showed computationally-complex Variance-of-Laplacian quality assessment accounts for \(\approx\)66\% of the additional processing overhead. 

Fig.~\ref{fig:R2-2} shows the per-image application latency observed in Experiment~1. Each experimental image comprised approximately 55 fragments on average. The resulting average application latency was \(21.23\)~ms, with 90th- and 95th-percentile values of \(32.76\)~ms and \(40.38\)~ms, respectively. Concurrent fragment processing and transmission allow the latency of these fragments to overlap rather than accumulate sequentially.  Thus, the proposed middleware enables an image to progress from transmission at \(U_S\), through reconstruction and IQA, to completion of object detection at \(U_{Yv8}\) in \(21.23\)~ms on average, with 95\% of images completing the end-to-end perception pipeline within \(40.38\)~ms. Communication and middleware throughput observed in Experiment 1 averaged \(4.24\)~Mbps and \(2.22\)~Mbps, respectively, over experiment duration. The reduction in observed middleware throughput is consistent with increase in latency from the additional reconstruction and IQA processing time. 


\begin{figure}[h]
    \centering
    \includegraphics[width=0.9\linewidth]{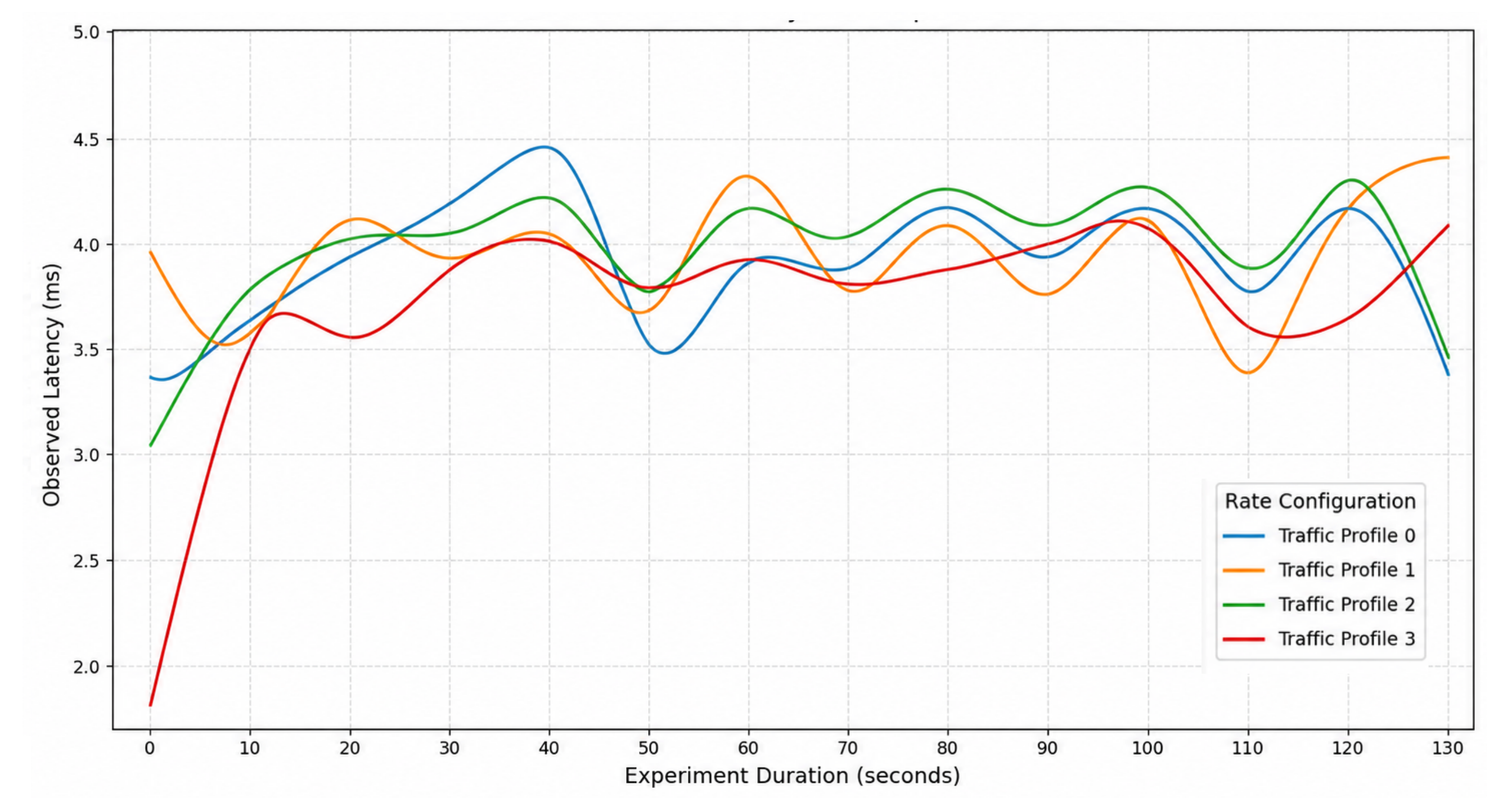}
    \caption{Latency Observed for High-Priority Image Packets under Varying Background Traffic Conditions.}
    \label{fig:R3}
\end{figure}

Fig.~\ref{fig:R3} shows the communication latency of high-priority image packets under increasing lower-priority background traffic. Traffic Profiles 0--3 correspond to \(U'_S\) transmitting similarly sized lower-priority packets at \(1\times\), \(2\times\), \(3\times\), and \(4\times\) the transmission rate of \(U_S\), respectively, while \(U_S\) transmitted with a higher-priority. Despite increasing background traffic from \(1\times\) to \(4\times\), the tail latency of high-priority image packets increased only marginally, by approximately \(1\)~ms across the evaluated traffic profiles. These results provide an initial validation that the broker's priority-aware scheduler can identify and preferentially service high-priority perception traffic under increasing load. A more comprehensive evaluation, including the resulting latency and potential starvation of competing lower-priority traffic, is left for future work.

Table~\ref{tab:T2} illustrates the middleware's IQA capability as demonstrated through impact on YOLOv8n's object detection performance. Results are shown for the \textit{car} class, which is sufficient to evaluate the impact of degraded perception data on the target application. Overall detection performance is relatively modest. This is consistent with the class imbalance and baseline performance reported for the AU-AIR dataset~\cite{bozcan2020air}. With IQA disabled, introducing 20 blurred images reduces precision, recall, mAP@50, and mAP@50--95 by 12.22\%, 14.11\%, 18.14\%, and 20.56\%, respectively. With increasing degradation, recall decreases from 0.801 to 0.548 and mAP@50 from 0.645 to 0.479, while precision stabilizes near 0.63 after its initial decline. This suggests that further degradation primarily increases missed detections rather than false positives. With IQA enabled, the corresponding reductions at 20 blurred images are limited to 3.37\%, 1.62\%, 3.41\%, and 6.60\%. Metrics subsequently increase at 40 and 60 blurred images because filtering changes the composition of images reaching YOLOv8n, creating a \textit{selection effect} rather than improving the detector itself. Thus, IQA prevents unsuitable perception data from reaching YOLOv8n and substantially mitigates degradation in object detection performance.

\begin{table}[t]
\centering
\caption{Impact of IQA on Object Detection Performance.}
\label{tab:T2}
\footnotesize

\begin{tabular}{|c|c|c|c|c|}
\hline
\textbf{Images Blurred} & 
\textbf{Precision} & 
\textbf{Recall} & 
\textbf{mAP@50} & 
\textbf{mAP@50--95} \\
\hline

\multicolumn{5}{|l|}{\textit{Image Quality Assessment: Off}} \\
\hline
0  & 0.712 & 0.801 & 0.645 & 0.394 \\
\hline
20 & 0.625 & 0.688 & 0.528 & 0.313 \\
\hline
40 & 0.628 & 0.624 & 0.517 & 0.308 \\
\hline
60 & 0.632 & 0.548 & 0.479 & 0.281 \\
\hline

\multicolumn{5}{|l|}{\textit{Image Quality Assessment: On}} \\
\hline
0  & 0.712 & 0.801 & 0.645 & 0.394 \\
\hline
20 & 0.688 & 0.788 & 0.623 & 0.368 \\
\hline
40 & 0.715 & 0.838 & 0.719 & 0.428 \\
\hline
60 & 0.739 & 0.887 & 0.807 & 0.472 \\
\hline

\end{tabular}
\end{table}

\section{Conclusion and Future Work} \label{conclusion}

In this work, we implemented and validated an initial version of a perception-aware communication middleware for distributed perception in heterogeneous UAV swarms. The middleware achieves low end-to-end communication and application latency while substantially improving throughput over a lightweight UDP broker baseline. Priority-aware scheduling maintains the latency of perception traffic under increasing background load, while middleware-level image quality assessment mitigates degradation in object detection performance caused by low-quality sensing data. Collectively, these results demonstrate the value of integrating perception awareness directly into the communication middleware. Future work will focus on iterative improvements and broader validation of the middleware. Thread management will be optimized to reduce tail latency, and priority-aware scheduling will be evaluated more extensively, particularly its impact on lower-priority traffic. The overarching goal is to package and release the middleware as an open-source platform for implementing and evaluating distributed perception applications in heterogeneous UAV swarms.

\bibliographystyle{IEEEtran}
\bibliography{refList}

\end{document}